\documentclass[11pt]{article}

\usepackage{amsmath}

\begin{document}

\title{E.M. Ovsiyuk, K.V. Kazmerchuk \\
Quasi-plane waves for spin 1 field in Lobachevsky \\ space
and a generalized
 helicity operator }

\maketitle

\date{}

\begin{abstract}
Spin 1  particle is investigated in  3-dimensional curved space of
constant  negative curvature. An extended helicity operator is
defined and the variables are separated in a tetrad-based
10-dimensional Duffin--Kemmer equation in quasi cartesian
 coordinates. The problem is solved exactly in
hypergeometric functions, the quantum states  are determined by three
 quantum numbers.
It is shown that Lobachevsky geometry acts effectively as a medium with
simple reflecting properties.
 Transition to a massless
case of electromagnetic field is performed.

\end{abstract}

This article continues a series of previous papers on constricting exact solutions
of the  wave equations   for fields of different spins   on the geometrical background of 3-dimensional spatial models
of constant positive or negative curvature [1--4].
Present paper is devoted to the (massive and massless) case of spin 1 field in  the
 Lobachevsky spacial model, the treatment is based on matrix
Duffin--Kemmer formalism applied in special quasi-cartesian coordinate of the Lobachevsky space.

In Lobachevsky space-time, let us use quasi-cartesian coordinates and
corresponding tetrad
$$
x^{a} = (t , x, y, z) \; , \qquad dS^{2}= dt^{2} - e^{-2z} (
dx^{2} + dy^{2} ) - dz^{2} \; ,
$$
$$
e_{(a)}^{\beta} = \left | \begin{array}{cccc}
1 & 0 & 0 & 0 \\
0 & e^{z} & 0 & 0 \\
0 & 0 & e^{z} & 0 \\
0 & 0 & 0 & 1
\end{array} \right | \; , \qquad
 e_{(a) \beta} = \left |
\begin{array}{llll}
1 & 0 & 0 & 0 \\
0 & -e^{-z} & 0 & 0 \\
0 & 0 & - e^{-z} & 0 \\
0 & 0 & 0 & - 1
\end{array} \right | \; .
\eqno(1)
$$

Christoffel symbols are $ \Gamma^{0}_{\beta
\sigma} = 0 \; , \; \Gamma^{i}_{00} = 0 \; , \; \Gamma^{i}_{0j} =
0$, and
$$
 x^{i}=(x,y,z)\; , \qquad
\Gamma^{i}_{\;\;jk } =  {1 \over 2} \; g^{i l} \; ( -\partial _{ l
} \; g_{jk} +
\partial _{j } \; g_{lk } + \partial_{k} \; g_{lj } ) \; ,
$$
$$
\Gamma^{x}_{\;\; jk} = \left | \begin{array}{ccc}
0 & 0 & -1 \\
0 & 0 & 0 \\
-1 & 0 & 0
\end{array} \right |  ,\;
 \Gamma^{y}_{\;\; jk} = \left | \begin{array}{ccc}
0 & 0 & 0 \\
0 & 0 & -1 \\
0 & -1 & 0
\end{array} \right |  ,\;
\Gamma^{z}_{\;\; jk} = \left | \begin{array}{ccc}
e^{-2z} & 0 & 0 \\
0 & e^{-2z} & 0 \\
0 & 0 & 0
\end{array} \right | \; .
$$

\noindent The Ricci rotations coefficients are
$$
\gamma_{01 1}= \gamma_{02 1}= \gamma_{03 1} = 0 \, , \;
\gamma_{012}= \gamma_{022}= \gamma_{032} = 0 \, , \; \gamma_{01
3}= \gamma_{02 3}= \gamma_{03 3} = 0 \, ,
$$
$$
\gamma_{23 1}  = 0 \, , \; \gamma_{31 1}  = -1\, , \;
\gamma_{12 1}= 0 \, , \; \gamma_{23 2} = 1 \, , \; \gamma_{31
2} = 0 \, , \;
 \gamma_{12 2} = 0 \, ,
$$
$$
\gamma_{23 3} = 0 \; , \; \gamma_{31 3}  = 0\; , \; \gamma_{12 3}
= 0 \; .
$$

Duffin--Kemmer equation [5]
$$
\left [ i\beta^{c} \left  (\ e_{(c)}^{\beta}  \partial_{\beta} +
{1 \over 2 } J^{ab} \gamma_{abc}  \right ) - M \right ] \Psi = 0
\eqno(2)
$$

\noindent in the above tetrsd takes the form
$$
\left [  +i \beta^{0}  {\partial \over \partial t} + i \beta ^{1}
e^{z} {\partial \over \partial x} + i \beta^{2}  e^{z}  {\partial
\over \partial y}
 + i \left ( \beta^{3}  {\partial \over \partial z}
-  \beta^{1}  J^{31}   + \beta^{2}  J^{23} \right )    - M \right
]  \Psi =0 \; .\eqno(3)
$$

We will search solutions in the form of quasi-plane waves
$$
\Psi = e^{-i \epsilon  t } \; e^{i a x} \; e^{ib  y} \left |
\begin{array}{c}   \Phi_{0}(z)  \\ \Phi_{j} (z) \\  E_{j} (z) \\
H_{j} (z)
\end{array} \right |,
\eqno(4)
$$

\noindent eq.  (3) gives
$$
\left [  \epsilon  \beta^{0}    -a  \beta ^{1}  e^{z} - b
\beta^{2}  e^{z}
 + i \left ( \beta^{3}  {\partial \over \partial z}
 -\beta^{1}  J^{31}   + \beta^{2}  J^{23} \right )    - M \right ]  \Psi =0\,.
\eqno(5)
$$

Below we will use the cyclic basis for Duffin--Kemmer matrices, then
a third projection of the spin is diagonal matrix
$$
\beta^{0} = \left | \begin{array}{rrrr}
 0       &   0        &  0  &  0 \\
 0  &  0       &  i  & 0  \\
  0  &   -i       &   0  & 0\\
   0  &  0       &   0  & 0
\end {array} \right |
, \qquad \beta^{i} = \left |
\begin{array}{rrrrr}
  0       &  0       &    e_{i}  & 0       \\
    0   &  0       &   0      & \tau_{i} \\
   -e_{i}^{+}  &  0       &   0      & 0       \\
   0       &  -\tau_{i}&   0      & 0
\end {array} \right | ;
$$
$$
e_{1} = {1 \over \sqrt{2}} ( -i, \; 0  , \; i )\; , \qquad e_{2} =
{1 \over \sqrt{2}} ( 1 , \; 0  , \;  1 )\; , \qquad e_{3} = ( 0 ,
i  , 0)\; ;
$$
$$
\tau_{1} = {1 \over \sqrt{2}} \left |  \begin{array}{ccc} 0  &  1
&  0  \\ 1 &  0  &  1  \\ 0  &  1  &  0
\end{array} \right | ,\;
\tau_{2}= {1 \over \sqrt{2}} \left |
\begin{array}{ccc} 0  &  -i  &  0  \\ i & 0  &  -i  \\ 0  &  i  &
0
\end{array} \right | ,\;
  \tau_{3} =   \left |
\begin{array}{rrr} 1  &  0  &  0  \\ 0  &  0  &  0   \\ 0  &  0
&  -1, \end{array} \right |  =  s_{3}\; ;
$$
$$
 J^{12} =  \beta^{1} \beta^{2} -  \beta^{2} \beta^{1}
=
  -i \left | \begin{array}{cccc}
0 & 0  &  0 & 0  \\
0 &   \tau_{3}  & 0 & 0 \\
0 & 0 &  \tau_{3} & 0 \\
0 & 0 & 0 &  \tau_{3}
\end{array} \right | = -iS_{3} \; ,
$$
$$
 J^{13} =  \beta^{1} \beta^{3} -  \beta^{3} \beta^{1}
=
 i \left | \begin{array}{cccc}
0 & 0  &  0 & 0  \\
0 &   \tau_{2}  & 0 & 0 \\
0 & 0 &  \tau_{2} & 0 \\
0 & 0 & 0 &  \tau_{2}
\end{array} \right | = iS_{2}\;,
$$
$$
 J^{23} =  \beta^{2} \beta^{3} -  \beta^{3} \beta^{2}
=
  -i \left | \begin{array}{cccc}
0 & 0  &  0 & 0  \\
0 &   \tau_{1}  & 0 & 0 \\
0 & 0 &  \tau_{1} & 0 \\
0 & 0 & 0 &  \tau_{1}
\end{array} \right | = -iS_{1}\; ;
$$
$$
-\beta^{1} J^{31} + \beta^{2} J^{23} = i \left |
\begin{array}{cccc}
0 & 0 & (e_{1}\tau_{2} -e_{2}\tau_{1}) & 0 \\
0 & 0 & 0  & (\tau_{1} \tau_{2} - \tau_{2} \tau_{1}) \\
0 & 0 & 0 & 0
\\
0 & - (\tau_{1} \tau_{2} - \tau_{2} \tau_{1})& 0 & 0
\end{array} \right |.
$$

Eq.  (5) in  block form reads
$$
\left [ \epsilon \left | \begin{array}{rrrr}
 0       &   0        &  0  &  0 \\
 0  &  0       &  i  & 0  \\
  0  &   -i       &   0  & 0\\
   0  &  0       &   0  & 0
\end {array} \right | -a e^{z}
\left |
\begin{array}{rrrrr}
  0       &  0       &    e_{1}  & 0       \\
    0   &  0       &   0      & \tau_{1} \\
   -e_{1}^{+}  &  0       &   0      & 0       \\
   0       &  -\tau_{1}&   0      & 0
\end {array} \right |-\right.
$$
$$\left.
-b e^{z} \left |
\begin{array}{rrrrr}
  0       &  0       &    e_{2}  & 0       \\
    0   &  0       &   0      & \tau_{2} \\
   -e_{2}^{+}  &  0       &   0      & 0       \\
   0       &  -\tau_{2}&   0      & 0
\end {array} \right |  
 + i  ( \;\;  \left |
\begin{array}{rrrrr}
  0       &  0       &    e_{3}  & 0       \\
    0   &  0       &   0      & \tau_{3} \\
   -e_{3}^{+}  &  0       &   0      & 0       \\
   0       &  -\tau_{3}&   0      & 0
\end {array} \right | {\partial \over \partial  z} + \right.
$$
$$
\left. \left.  + i \left | \begin{array}{cccc}
0 & 0 & (e_{1}\tau_{2} -e_{2}\tau_{1}) & 0 \\
0 & 0 & 0  & (\tau_{1} \tau_{2} - \tau_{2} \tau_{1}) \\
0 & 0 & 0 & 0
\\
0 & - (\tau_{1} \tau_{2} - \tau_{2} \tau_{1})& 0 & 0
\end{array} \right | \;\; \right )
 - M \right ]
\left | \begin{array}{c}   \Phi_{0}  \\ \Phi_{j}  \\  E_{j}
 \\  H_{j}
\end{array} \right |=0,
\eqno(6a)
$$

\noindent which is equivalent to
$$
-a\,e^{z}\,e_{1}\,\vec{E}-b\,e^{z}\,e_{2}\,\vec{E}+i\,e_{3}\,{\partial
\vec{E}\over \partial
z}-(e_{1}\tau_{2}-e_{2}\,\tau_{1})\,\vec{E}-M\,\Phi_{0}=0\,,
$$
$$
\epsilon\,I\,\vec{E}-a\,e^{z}\,\tau_{1}\,\vec{H}-b\,e^{z}\,\tau_{2}\,\vec{H}+i\,\tau_{3}\,{\partial
\vec{H}\over \partial
z}-(\tau_{1}\tau_{2}-\tau_{2}\,\tau_{1})\,\vec{H}-M\,\vec{\Phi}=0\,,
$$
$$
-\epsilon\,I\,\vec{\Phi}+a\,e^{z}\,e_{1}^{+}\,\Phi_{0}+b\,e^{z}\,e_{2}^{+}\,\Phi_{0}-i\,e_{3}^{+}\,{\partial
\Phi_{0}\over \partial z}-M\,\vec{E}=0\,,
$$
$$
a\,e^{z}\,\tau_{1}\,\vec{\Phi}+
b\,e^{z}\,\tau_{2}\,\vec{\Phi}-i\,\tau_{3}\,{\partial
\vec{\Phi}\over \partial z}+(\tau_{1}\tau_{2}-
\tau_{2}\,\tau_{1})\,\vec{\Phi}-M\,\vec{H}=0 \; . \eqno(6b)
$$

\noindent In explicit form, 10 equations are
 (let  $\gamma = 1 /
\sqrt{2}$)
$$
i\gamma \,a\,e^{z}\,(E_{1}-E_{3})- \gamma
\,b\,e^{z}\,(E_{1}+E_{3})-{d E_{2}\over
dz}+2\,E_{2}-M\,\Phi_{0}=0\,, \eqno(7a)
$$

$$
i\,\epsilon\,E_{1}- \gamma \,a\,e^{z}\,H_{2}+ i \gamma
\,b\,e^{z}\,H_{2}+i\,{d H_{1}\over dz}-i\,H_{1}-M\,\Phi_{1}=0\,,
$$
$$
i\,\epsilon\,E_{2}- \gamma a\,e^{z}\,(H_{1}+H_{3})-
i\gamma\,b\,e^{z}\,(H_{1}-H_{3})-M\,\Phi_{2}=0\,,
$$
$$
i\,\epsilon\,E_{3}- \gamma a\,e^{z}\,H_{2}- i\gamma
\,b\,e^{z}\,H_{2}-i\,{d H_{3}\over dz}+i\,H_{3}-M\,\Phi_{3}=0\,,
\eqno(7b)
$$

$$
-i\,\epsilon\,\Phi_{1}+ i\gamma \,a\,e^{z}\,\Phi_{0}+ \gamma
b\,e^{z}\,\Phi_{0}-M\,E_{1}=0\,,
$$
$$
-i\,\epsilon\,\Phi_{2}-{d \Phi_{0}\over dz}-M\,E_{2}=0\,,
$$
$$
-i\,\epsilon\,\Phi_{3}- i\gamma \,a\,e^{z}\,\Phi_{0}+ \gamma
b\,e^{z}\,\Phi_{0}-M\,E_{3}=0\,, \eqno(7c)
$$

$$
\gamma a\,e^{z}\,\Phi_{2}- i\gamma \,b\,e^{z}\,\Phi_{2}-i\,{d
\Phi_{1}\over dz}+i\,\Phi_{1}-M\,H_{1}=0\,,
$$
$$
\gamma a\,e^{z}\,(\Phi_{1}+\Phi_{3})+ i \gamma
\,b\,e^{z}\,(\Phi_{1}-\Phi_{3})-M\,H_{2}=0\,,
$$
$$
\gamma a\,e^{z}\,\Phi_{2}+ i \gamma \,b\,e^{z}\,\Phi_{2}+i\,{d
\Phi_{3}\over dz}-i\,\Phi_{3}-M\,H_{3}=0\,. \eqno(7d)
$$

\noindent After evident regrouping they look
$$
\gamma (i a  - b ) e^{z}\,E_{1} - \gamma (ia +b) e^{z} E_{3} - ({d
\over dz} -2) \,E_{2}-M\,\Phi_{0}=0\,, \eqno(8a)
$$
$$
i\,\epsilon\,E_{1}- \gamma  (a -ib) \,e^{z}\,H_{2}+
 i\, ({d \over dz}- 1) \,H_{1}-M\,\Phi_{1}=0\,,
$$
$$
i\,\epsilon\,E_{2}-  \gamma (a +ib) \,e^{z}\,H_{1} - \gamma(a -
ib)  \,e^{z}\, H_{3} -M\,\Phi_{2}=0\,,
$$
$$
i\,\epsilon\,E_{3}- \gamma ( a +ib) \,e^{z}\,H_{2} -i (\,{d \over
dz}-1) \,H_{3}-M\,\Phi_{3}=0\,, \eqno(8b)
$$

$$
-i\,\epsilon\,\Phi_{1}+ \gamma\;  (b+ ia) \,e^{z}\,\Phi_{0}
-M\,E_{1}=0\,,
$$
$$
-i\,\epsilon\,\Phi_{2} - {d \Phi_{0}\over dz}-M\,E_{2}=0\,,
$$
$$
-i\,\epsilon\,\Phi_{3}  + \gamma (b -ia)
\,e^{z}\,\Phi_{0}-M\,E_{3}=0\,, \eqno(8c)
$$

$$
\gamma (a -ib) \,e^{z}\,\Phi_{2}  - i ({d \over dz} -1)
\,\Phi_{1}-M\,H_{1}=0\,,
$$
$$
\gamma ( a+ib) \,e^{z}\, \Phi_{1} +  \gamma (a-ib) \,e^{z}\;
\Phi_{3}-M\,H_{2}=0\,,
$$
$$
\gamma (a+ib) \,e^{z}\,\Phi_{2}  + i\,({d \over dz}-1) \,\Phi_{3}
- M\,H_{3}=0\,. \eqno(8d)
$$

To find  explicit solutions of the system,  it is convenient to
diagonalize  additionally a generalized helicity operator (it explicit form
can be proposed by analogy reasons with the case of similar operator in Minkowski space
-- after that it is  the matter of simple calculation to verify that it commutes with the
wave operator\footnote{To avoid misunderstanding, it should be noted that  below we use an helicity operator
which differs in the facto $i$ from  usual, so its eigenvalues are imaginary or zero.}):
$$
\Psi_{0}(z)  = (\Phi_{0} (z),\Phi_{j}(z) , E_{j}(z) , H_{j}(z)
)\,,
$$
$$
\left [
    a e^{z} S^{1}  +  b e^{z} S^{2}
-i ( S^{3} {d \over dz} -  S^{1}  J^{31}   + S^{2}  J^{23} ) \right ]
\Psi_{z} = \sigma \Psi (z)   \; ,
$$
$$
-  S^{1}  J^{31}   + S^{2}  J^{23}= - S^{3} \eqno(9a)
$$
\noindent in block form it reads
$$
\left [ a e^{z} \left | \begin{array}{cccc}
0 & 0  &  0 & 0  \\
0 &   \tau_{1}  & 0 & 0 \\
0 & 0 &  \tau_{1} & 0 \\
0 & 0 & 0 &  \tau_{1}
\end{array} \right |
+b e^{z} \left | \begin{array}{cccc}
0 & 0  &  0 & 0  \\
0 &   \tau_{2}  & 0 & 0 \\
0 & 0 &  \tau_{2} & 0 \\
0 & 0 & 0 &  \tau_{2}
\end{array} \right | - \right.
$$
$$
\left. -  i   \left | \begin{array}{cccc}
0 & 0  &  0 & 0  \\
0 &   \tau_{3}  & 0 & 0 \\
0 & 0 &  \tau_{3} & 0 \\
0 & 0 & 0 &  \tau_{3}
\end{array} \right | ({d \over d  z} -1) \right ]
\left | \begin{array}{c}   \Phi_{0}(z)  \\ \Phi_{j} (z) \\  E_{j}
(z) \\  H_{j} (z)
\end{array} \right |=\sigma \left | \begin{array}{c}   \Phi_{0}(z)  \\ \Phi_{j} (z) \\  E_{j} (z) \\  H_{j} (z)
\end{array} \right | \,,
\eqno(9b)
$$

\noindent so that
$$
0 =\sigma\,\Phi_{0}\,,
$$
$$
a\,e^{z}\,\tau_{1}\,\vec{\Phi}+b\,e^{z}\,\tau_{2}\,\vec{\Phi}-
i\,\tau_{3}\,({d \over d z} -1)  \vec{\Phi}
=\sigma\,\vec{\Phi}\,,
$$
$$
a\,e^{z}\,\tau_{1}\,\vec{E}+b\,e^{z}\,\tau_{2}\,\vec{E}-i\,\tau_{3}\,{d
\vec{E}\over d z}=\sigma\,\vec{E}\,,
$$
$$
a\,e^{z}\,\tau_{1}\,\vec{H}+
b\,e^{z}\,\tau_{2}\,\vec{H}-i\,\tau_{3}\,( {d \over d z} -1)
\vec{H} =\sigma\,\vec{H} \; . \eqno(9c)
$$

\noindent After simple calculations we get 10 equations
$$
0  =\sigma\,\Phi_{0}\,,
$$
$$
\gamma \,(a-ib) \,e^{z}\,\Phi_{2}-
 i\,( {d \over dz} -1 ) \Phi_{1}  =\sigma\,\Phi_{1}\,,
$$
$$
 \gamma (a+ib) \,e^{z}\, \Phi_{1} +
i\gamma\,( a - ib)\,e^{z}\,\Phi_{3} =\sigma\,\Phi_{2}\,,
$$
$$
\gamma (a+ib) \,e^{z}\,\Phi_{2}+
  i\,( {d \over dz}-1)  \Phi_{3}  =\sigma\,\Phi_{3}\,,
$$
$$
\gamma \,(a-ib) \,e^{z}\,E_{2} - i\,{d E_{1}\over
dz}=\sigma\,E_{1}\,,
$$
$$
 \gamma (a+ib) \,e^{z}\,E_{1} +
\gamma\,(a-ib) \,e^{z}\,E_{3} =\sigma\,E_{2}\,,
$$
$$
\gamma (a+ib) \,e^{z}\,E_{2} + i\,{d E_{3}\over
dz}=\sigma\,E_{3}\,,
$$
$$
\gamma \,(a-ib) \,e^{z}\,H_{2}- i\,({d \over dz}-1)  H_{1}
=\sigma\,H_{1}\,,
$$
$$
 \gamma (a+ib) \,e^{z}\,H_{1}+
\gamma\,(a-i b) \,e^{z}\,H_{3}=\sigma\,H_{2}\,,
$$
$$
\gamma (a+ib) \,e^{z}\,H_{2} +i\,({d \over dz}-1) H_{3}
=\sigma\,H_{3}\, .
$$

\noindent It is more convenient to rewrite them in the form
$$
0  =\sigma\,\Phi_{0}\,, \eqno(10a)
$$
$$
-
 i\,( {d \over dz} -1 ) \Phi_{1}  =\sigma\,\Phi_{1} - \gamma \,(a-ib) \,e^{z}\,\Phi_{2} \,,
$$
$$
 \gamma (a+ib) \,e^{z}\, \Phi_{1} +
\gamma\,( a - ib)\,e^{z}\,\Phi_{3} =\sigma\,\Phi_{2}\,,
$$
$$
+
  i\,( {d \over dz}-1)  \Phi_{3}  =\sigma\,\Phi_{3}- \gamma (a+ib) \,e^{z}\,\Phi_{2} \,,
\eqno(10b)
$$
$$
 - i\,{d E_{1}\over dz}=\sigma\,E_{1}- \gamma \,(a-ib) \,e^{z}\,E_{2}\,,
$$
$$
 \gamma (a+ib) \,e^{z}\,E_{1} +
\gamma\,(a-ib) \,e^{z}\,E_{3} =\sigma\,E_{2}\,,
$$
$$
+ i\,{d E_{3}\over dz}=\sigma\,E_{3}- \gamma (a+ib) \,e^{z}\,E_{2}
\,, \eqno(10c)
$$
$$
- i\,({d \over dz}-1)  H_{1} =\sigma\,H_{1} - \gamma \,(a-ib)
\,e^{z}\,H_{2} \,,
$$
$$
 \gamma (a+ib) \,e^{z}\,H_{1}+
\gamma\,(a-i b) \,e^{z}\,H_{3}=\sigma\,H_{2}\,,
$$
$$
 +i\,({d \over dz}-1) H_{3} =\sigma\,H_{3}- \gamma (a+ib) \,e^{z}\,H_{2} \,.
\eqno(10d)
$$

Note that we see three  similar groups of equations for $\Phi_{j}, E_{j}, H_{i}$  respectively.
Let us examine one of them, foe instance
$$
\gamma \,(a-ib) \,e^{z}\,H_{2}- i\,({d \over dz}-1)  H_{1}
=\sigma\,H_{1}\,,
$$
$$
 \gamma (a+ib) \,e^{z}\,H_{1}+
\gamma\,(a-i b) \,e^{z}\,H_{3}=\sigma\,H_{2}\,,
$$
$$
\gamma (a+ib) \,e^{z}\,H_{2} +i\,({d \over dz}-1) H_{3}
=\sigma\,H_{3}\,. \eqno(11)
$$

The case  $\sigma =0$ reduces to
 (functions $H_{1}$ and  $H_{3}$ turn to be proportional to each other)
 $$
  (a+ib) H_{1} = -  (a-i b) H_{3}  \,,
$$
$$
H_{2} = +{ie^{-z } \over \gamma (a-ib)} ({d \over dz} - 1) H_{1} =
-{ie^{-z } \over \gamma (a+ib)} ({d \over dz} - 1) H_{3} \; .
\eqno(12)
$$

When  $\sigma \neq 0$, one  can exclude  $H_{2}$, then the first ant the third equations
 provide us  with the linear differential  system  for  $H_{1}$ and $H_{3}$:
$$
(a-ib) \,e^{2z}\,\left [  (a+ib) H_{1}+ (a-i b)  H_{3}\right ]- i
\sigma \,({d \over dz}-1)  H_{1} = 2\,\sigma^{2}\,H_{1}\,,
$$
$$
 (a+ib) \,e^{2z}\,\left [  (a+ib) H_{1}+
(a-i b)  H_{3}\right] +i \sigma \,({d \over dz}-1) H_{3} =2 \,
\sigma^{2}  \,H_{3}\, .
$$
$$
\eqno(13)
$$

\noindent One may observe the symmetry in (13):
$$
H_{1} \Longrightarrow H_{3} \; , \qquad \sigma \Longrightarrow -
\sigma \; ,
 \qquad b \Longrightarrow - b \; .
$$

\noindent
From (13) we get
$$
H_{3}={2i\,\sigma\over e^{2z}\,(a-i\,b)^{2}}\,\left({d\over
dz}-1-i\,\sigma\right)\,H_{1}-{a^{2}+b^{2}\over
(a-i\,b)^{2}}\,H_{1}\,,
$$
$$
{d^{2}H_{1}\over dz^{2}}-4\,{dH_{1}\over dz}- \left [
e^{2z}\,(a^{2}+b^{2})-4-(i+\sigma)^{2} \right ]\,H_{1}=0\, ,
\eqno(14a)
$$

\noindent and  the second symmetrical variant
$$
H_{1}={- 2i\,\sigma\over  e^{2z}\, (a+i\,b)^{2}}\,\left({d\over
dz}-1+i\,\sigma\right)\,H_{3}-{a^{2}+b^{2}\over
(a+i\,b)^{2}}\,H_{3}\,,
$$
$$
{d^{2}H_{3}\over dz^{2}}-4\,{dH_{3}\over dz}- \left [
e^{2z}\,(a^{2}+b^{2})-4-(i-\sigma)^{2} \right ]\,H_{3}=0\, .
\eqno(14b)
$$

\noindent  Let us translate  eqs.  $(14a)$ to other variable
$$
i \sqrt{a^{2} + b^{2} } e^{z} = Z \;, \qquad {d \over dz} = Z {d
\over dZ},
$$

\noindent then the second order  equations reads
$$
\left [ Z^{2} {d^{2} \over dZ^{2} } -3Z {d \over dZ } + Z^{2} + 4
- (1-i\sigma) ^{2}  \right ]  H_{1} = 0 \eqno(15a)
$$

\noindent or differently
$
H_{1} (Z) = Z^{2}  h_{1}(Z)
$:
$$
\left [ { d^{2} \over d Z^{2} } + {1 \over Z} {d \over d Z}  +1 -
{ (1-i\sigma)^{2} \over Z^{2} } \right ] h_{1} = 0 \; ,
\eqno(15b)
$$

\noindent which is the Bessel equations with solutions
$$
h_{1}= J_{\pm \mu}  ( Z)\;, \qquad \mu = 1 -i\sigma \; .
\eqno(15c)
$$

\noindent Expression   $h_{3}$ $(14a)$ is defined by
$$
Z^{2} h_{3} = { a +ib \over a-ib} \left [ -2i\sigma (Z {d\over d
Z} + 1 - i \sigma  ) - Z^{2} \right]  h_{1} \eqno(15d)
$$

Similar formulas can be produced for $(14b)$:
$$
H_{3} = e^{2Z} h_{3}
$$

\noindent and $$
\left [ { d^{2} \over d Z^{2} } + {1 \over Z} {d \over d Z}  +1 -
{ (1+i\sigma)^{2} \over Z^{2} } \right ] h_{3} = 0 \; ;
\eqno(16a)
$$
$$
h_{3}= J_{\pm \nu}  ( Z)\;, \qquad \nu = 1 +i\sigma \; ;
\eqno(16b)
$$
$$
Z^{2} h_{1} = { a -ib \over a+ ib} \left [ + 2i\sigma (Z {d\over d
Z} + 1 + i \sigma ) - Z^{2} \right]  h_{3}  \; .
\eqno(16c)
$$

Now we are to joint together the main equations   (8) and  equations  (11).

First, let us consider the case of non-zero $\sigma$. At this, from the very beginning, one must assume
   $\Phi_{0}= 0$. Eqs.   (8)  giv
$$
\gamma (i a  - b ) e^{z}\,E_{1} - \gamma (ia +b) e^{z} E_{3} - ({d
\over dz} -2) \,E_{2}=0\,, \eqno(17a)
$$
$$
i\,\epsilon\,E_{1} - \sigma\,H_{1} - M\,\Phi_{1}=0\,,
$$
$$
i\,\epsilon\,E_{2}  - \sigma\,H_{2}   - M\,\Phi_{2}=0\,,
$$
$$
i\,\epsilon\,E_{3}  - \sigma\,H_{3}  - M\,\Phi_{3}=0\,,
\eqno(17b)
$$
$$
-i\,\epsilon\,\Phi_{1} -M\,E_{1}=0\,, \qquad
-i\,\epsilon\,\Phi_{2}  -M\,E_{2}=0\,, \qquad
-i\,\epsilon\,\Phi_{3}   -M\,E_{3}=0\,, \eqno(17c)
$$
$$
 \sigma\,\Phi_{1}  - M\,H_{1}=0\,, \qquad
\sigma\,\Phi_{2} -M\,H_{2}=0\,, \qquad
\sigma\,\Phi_{3} - M\,H_{3}=0\,. 
$$
$$
\eqno(17d)
$$

\noindent
On can exclude   $E_{j}$ from eq.
$$
\gamma (i a  - b ) e^{z}\,\Phi_{1} - \gamma (ia +b) e^{z} \Phi_{3}
- ({d \over dz} -2) \,\Phi_{2}=0\,. \eqno(18)
$$

This relation coincides with the Lorentz condition (when  $\Phi_{0}=0$).
Indeed, let us start with the Lorentz condition in tensor form
$$
\nabla_{\beta} \Phi^{\beta(cart)} =0 \qquad  \Longrightarrow
\qquad  \nabla_{\beta} ( e^{(b)\beta} \Phi_{(b)}^{cart}  =0\qquad
\Longrightarrow
$$
$$
 { \partial \Phi_{(b)^{cart} } \over \partial x^{\beta} }
   \;    e^{(b)\beta} +   \Phi_{(b)} ^{cart}  \nabla_{\beta}  e^{(b) \beta}   =0 \; ,
\eqno(19a)
$$

\noindent or
$$
 { \partial \Phi_{(b) }^{cart} \over \partial x^{\beta} }
   \;    e^{(b)\beta} +   \Phi_{(b)}^{cart}  { 1 \over \sqrt{-g}} {\partial \over \partial x^{\beta}}
\sqrt{-g}   e^{(b) \beta}   =0 \; . \eqno(19b)
$$

\noindent Allowing for relations  (1), eq. (19b) reduces to
$$
 { \partial \Phi_{(0) } ^{cart}\over \partial t }
 -e^{z}  { \partial \Phi_{(1) }^{cart} \over \partial x}
-e^{z}  { \partial \Phi_{(2) }^{cart} \over \partial y} - ( {
\partial   \over \partial z } - 2)  \Phi_{(3)}^{cart}  =0 \; ,
\eqno(19c)
$$

\noindent
or with the substitution
$$
(\Phi^{cart}_{a})  = e^{-i\epsilon t }  e^{iax} e^{iby} \left |
\begin{array}{c}
\Phi^{cart}_{0}(z)\\
\Phi^{cart}_{1}(z)\\
\Phi^{cart}_{2}(z)\\
\Phi^{cart}_{3}(z)
\end{array} \right |
$$

\noindent  we arrive at
$$
 -i \epsilon  \Phi_{(0) } ^{cart}
 - ia e^{z}   \Phi_{(1) }^{cart}
-ib e^{z}   \Phi_{(2) }^{cart} - ( { d    \over d  z } - 2)
\Phi_{(3)}^{cart}  =0 \; . \eqno(20a)
$$

\noindent
Translating the last relation  to the variables of cyclic basis
 $$
\Phi_{2} = \Phi^{cart}_{(3)}\; , \qquad \Phi_{3} - \Phi_{1} =
\sqrt{2}  \Phi_{(1)}^{cart}   \;, \qquad \Phi_{3} + \Phi_{1}
=\sqrt{2}i\;  \Phi_{(2)}^{cart}\;,
$$

\noindent we get
$$
 -i \epsilon  \Phi_{0 }
 - ia e^{z}   {\Phi_{3} - \Phi_{1} \over \sqrt{2} }
-ib e^{z}   {\Phi_{3} + \Phi_{1} \over \sqrt{2} i  } - ( { d
\over d  z } - 2)  \Phi_{2}  =0 \; ,
$$

\noindent that is
$$
 -i \epsilon  \Phi_{0 } + {ia -b \over \sqrt{2}} e^{z} \Phi_{1} -
{ia +b \over \sqrt{2}} e^{z} \Phi_{3} - ( { d    \over d  z } - 2)
\Phi_{2}  =0 \; . \eqno(20b)
$$

\noindent
Eq. (20b) when  $\Phi_{0}=0$ coincides with (18)
$$
\gamma (i a  - b ) e^{z}\,\Phi_{1} - \gamma (ia +b) e^{z} \Phi_{3}
- ({d \over dz} -2) \,\Phi_{2}=0\, .
$$

 Therefore, eq.  (18) (or$(17a)$) is  identity that is valid automatically  due to
 the structure of the wave equation for spin 1  field.
 Remaining 9 equations  $(17b,c,d)$ lead us to linear constrains  between
9 nontrivial constituents:
$$
i\,\epsilon\,E_{j} - \sigma\,H_{j} - M\,\Phi_{j}=0\,,
$$
$$
-i\,\epsilon\,\Phi_{j} -M\,E_{j}=0\,,
$$
$$
 \sigma\,\Phi_{j}  - M\,H_{j}=0\,,
\eqno(21a)
$$

\noindent that is
$$
\sigma = \pm \sqrt{\epsilon^{2}- M^{2} }\;, \qquad \Phi_{0}= 0\;,
$$
$$
E_{j} = -{i\epsilon \over M} \Phi_{j}\; , \qquad H_{j} = { \sigma
\over M} \Phi_{j}\;. \eqno(21b)
$$

Now, we are to proceed with the case $\sigma=0$. Eigenvalue helicity equations here are
$$
0  =0\,, \eqno(22a)
$$
$$
-  i\,( {d \over dz} -1 ) \Phi_{1}  = - \gamma \,(a-ib)
\,e^{z}\,\Phi_{2} \,,
$$
$$
 \gamma (a+ib) \,e^{z}\, \Phi_{1} +
\gamma\,( a - ib)\,e^{z}\,\Phi_{3} =0\,,
$$
$$
+
  i\,( {d \over dz}-1)  \Phi_{3}  =- \gamma (a+ib) \,e^{z}\,\Phi_{2} \,,
\eqno(22b)
$$
$$
 - i\,{d E_{1}\over dz}=- \gamma \,(a-ib) \,e^{z}\,E_{2}\,,
$$
$$
 \gamma (a+ib) \,e^{z}\,E_{1} +
\gamma\,(a-ib) \,e^{z}\,E_{3} =0\,,
$$
$$
+ i\,{d E_{3}\over dz}= - \gamma (a+ib) \,e^{z}\,E_{2} \,,
\eqno(22c)
$$
$$
- i\,({d \over dz}-1)  H_{1} = - \gamma \,(a-ib) \,e^{z}\,H_{2}
\,,
$$
$$
 \gamma (a+ib) \,e^{z}\,H_{1}+
\gamma\,(a-i b) \,e^{z}\,H_{3}=0 \,,
$$
$$
 +i\,({d \over dz}-1) H_{3} = - \gamma (a+ib) \,e^{z}\,H_{2} \,.
\eqno(22d)
$$

\noindent Taking into consideration  (22) from   (8) we get
$$
\gamma (i a  - b ) e^{z}\,E_{1} - \gamma (ia +b) e^{z} E_{3} - ({d
\over dz} -2) \,E_{2}-M\,\Phi_{0}=0\,, \eqno(23a)
$$
$$
i\,\epsilon\,E_{1}   -M\,\Phi_{1}=0\,,
$$
$$
i\,\epsilon\,E_{2}   - M\,\Phi_{2}=0\,,
$$
$$
i\,\epsilon\,E_{3}  -M\,\Phi_{3}=0\,, \eqno(23b)
$$
$$
-i\,\epsilon\,\Phi_{1}+ \gamma\;  (b+ ia) \,e^{z}\,\Phi_{0}
-M\,E_{1}=0\,,
$$
$$
-i\,\epsilon\,\Phi_{2} - {d \Phi_{0}\over dz}-M\,E_{2}=0\,,
$$
$$
-i\,\epsilon\,\Phi_{3}  + \gamma (b -ia)
\,e^{z}\,\Phi_{0}-M\,E_{3}=0\,, \eqno(23c)
$$
$$
 -M\,H_{1}=0\,, \qquad
-M\,H_{2}=0\,, \qquad
 - M\,H_{3}=0\,.
\eqno(23d)
$$

\noindent Excluding $E_{j}$ in   $(23c)$, we derive
$$
(\epsilon^{2} - M^{2} ) \Phi_{1} - \gamma \epsilon   (a -i b) \,e^{z}\,\Phi_{0}=0\,,
$$
$$
(\epsilon^{2} -M^{2}) \Phi_{2} - i \epsilon {d \over dz}  \Phi_{0} =0\,,
$$
$$
(\epsilon^{2} -M^{2} ) \Phi_{3}  + \gamma \epsilon (a+ ib )
\,e^{z}\,\Phi_{0}=0\,.
\eqno(24a)
$$

\noindent
Now let us recall  eqs.  (12)
$$
  (a+ib) \Phi_{1}+ (a-i b) \Phi_{3}= 0 \,,
$$
$$
\Phi_{2} = +{ie^{-z } \over \gamma (a-ib)} ({d \over dz} - 1) \Phi_{1} =
-{ie^{-z } \over \gamma (a+ib)} ({d \over dz} - 1) \Phi_{3} \; .
$$

\noindent
Thus, we  obtain equation for independent variables
$$
\Phi_{1} = + {\gamma \epsilon  (a -i b)  \over (\epsilon^{2} - M^{2} )  }   \,e^{z}\,\Phi_{0} \,,
$$
$$
 \Phi_{3} =
 - { \gamma \epsilon (a+ib) \over \epsilon^{2} -M^{2}}
 e^{z} \Phi_{0} \,,
$$
$$
\Phi_{2}=   {i \epsilon  \over (\epsilon^{2} -M^{2}) } {d \over dz}  \Phi_{0} \,.
\eqno(24c)
$$

\noindent
Substituting them into eq. $(23a)$
$$
\gamma (i a  - b ) e^{z}\,  \Phi _{1} - \gamma (ia +b) e^{z}  \Phi_{3} - ({d
\over dz} -2) \,  \Phi_{2}-  i \epsilon \Phi_{0}=0\,,
$$

\noindent
we arrive at
$$
\gamma (i a  - b ) e^{z}\,  {\gamma \epsilon  (a -i b)  \over (\epsilon^{2} - M^{2} )  }   \,e^{z}\,\Phi_{0}
 + \gamma (ia +b)   { \gamma \epsilon (a+ib) \over \epsilon^{2} -M^{2}}
 e^{z} \Phi_{0}  e^{z}  -
 $$
 $$
 - ({d
\over dz} -2) \,  {i \epsilon  \over (\epsilon^{2} -M^{2}) } {d \over dz}  \Phi_{0}  -  i \epsilon \Phi_{0}=0\,,
$$

\noindent
that is
$$
\left [   ({d
\over dz} -2) \,  {d \over dz}    + (\epsilon^{2} -M^{2})  - e^{2z} ( a^{2} + b^{2}) \right ]  \Phi_{0}=0\,.
\eqno(25)
$$

\noindent
It is of the type $(14a)$ and it can be  solved in Bessel functions.

In the end, one may mention that in massless case
instead of  (23) we would have
$$
0 =0\,,
$$
$$
i\,\epsilon\,E_{1}   =0\,, \qquad
i\,\epsilon\,E_{2}   =0\,, \qquad
i\,\epsilon\,E_{3}  =0\,,
$$
$$
-i\,\epsilon\,\Phi_{1}+ \gamma\;  (b+ ia) \,e^{z}\,\Phi_{0} =0\,,
$$
$$
-i\,\epsilon\,\Phi_{2} - {d \Phi_{0}\over dz}=0\,, \qquad
-i\,\epsilon\,\Phi_{3}  + \gamma (b -ia) \,e^{z}\,\Phi_{0}=0\,,
$$
$$
 H_{1}=0\,, \qquad
H_{2}=0\,, \qquad
 H_{3}=0\,.
\eqno(26)
$$

\noindent These equations describe gauge solution of gradient type,  with vanishing
electromagnetic tensor.

\vspace{3mm}

Authors are  grateful  to  V.M. Red'kov  for  moral support and  advices.

\end{document}